\documentclass[RNAAS]{aastex62}
\usepackage{amsmath}

\def\simlt{\mathrel{\hbox{\rlap{\hbox{\lower4pt\hbox{$\sim$}}}\hbox{$<$}}}}
\def\simgt{\mathrel{\hbox{\rlap{\hbox{\lower4pt\hbox{$\sim$}}}\hbox{$>$}}}}

\newcommand{\nimass}{M_\text{Ni56}}
\newcommand{\MWD}{M_{\rm{WD}}}

\begin{document}

\title{All known Type Ia supernovae models fail to reproduce the observed $t_0-\nimass$ correlation}

\correspondingauthor{Doron Kushnir}
\email{doron.kushnir@weizmann.ac.il}

\author{Amir Sharon}
\affiliation{Dept. of Particle Phys. \& Astrophys., Weizmann Institute of
Science, Rehovot 76100, Israel}

\author{Doron Kushnir}
\affiliation{Dept. of Particle Phys. \& Astrophys., Weizmann Institute of
Science, Rehovot 76100, Israel}

\begin{abstract}
Type Ia supernovae (SNe Ia) are likely the thermonuclear explosions of carbon-oxygen white-dwarf stars, but their progenitor systems remain elusive. A few theoretical scenarios for the progenitor systems have been suggested, which have been shown to agree with some observational properties of SNe Ia. However, several computational challenges prohibit a robust comparison to the observations. We focus on the observed $t_0-\nimass$ relation, where $t_0$ (the $\gamma$-rays' escape time from the ejecta) is positively correlated with $\nimass$ (the synthesized $^{56}$Ni mass). Comparing to the $t_0-\nimass$ relation bypasses the need for radiation transfer calculations, as the value of $t_0$ can be directly inferred from the ejecta. We show that all known SNe Ia models fail to reproduce the observed $t_0-\nimass$ correlation. 
\end{abstract}



\section{} 
Type Ia supernovae (SNe Ia) are likely the thermonuclear explosions of carbon-oxygen (CO) white-dwarf (WD) stars, but their progenitor systems remain elusive \citep[see][for a review]{Maoz2014}. A few theoretical scenarios for the progenitor systems have been suggested, including a Chandrasekhar-mass (Ch) or sub-Chandrasekhar-mass (sCh) WD that ignites due to some external interaction, and direct collision of WDs. Although these models have been shown to agree with some observational properties of SNe Ia, a few computational challenges prohibit a robust comparison to observations, such as the calculation of the propagating burning wave \citep[see, e.g.,][]{Ropke2017,KK2019} and accurate radiation transfer calculations \citep[see reviews, e.g.][]{Hillebrandt00,Noebauer2019}. 

\begin{figure}[h!]
\begin{center}
\includegraphics[scale=0.8,angle=0]{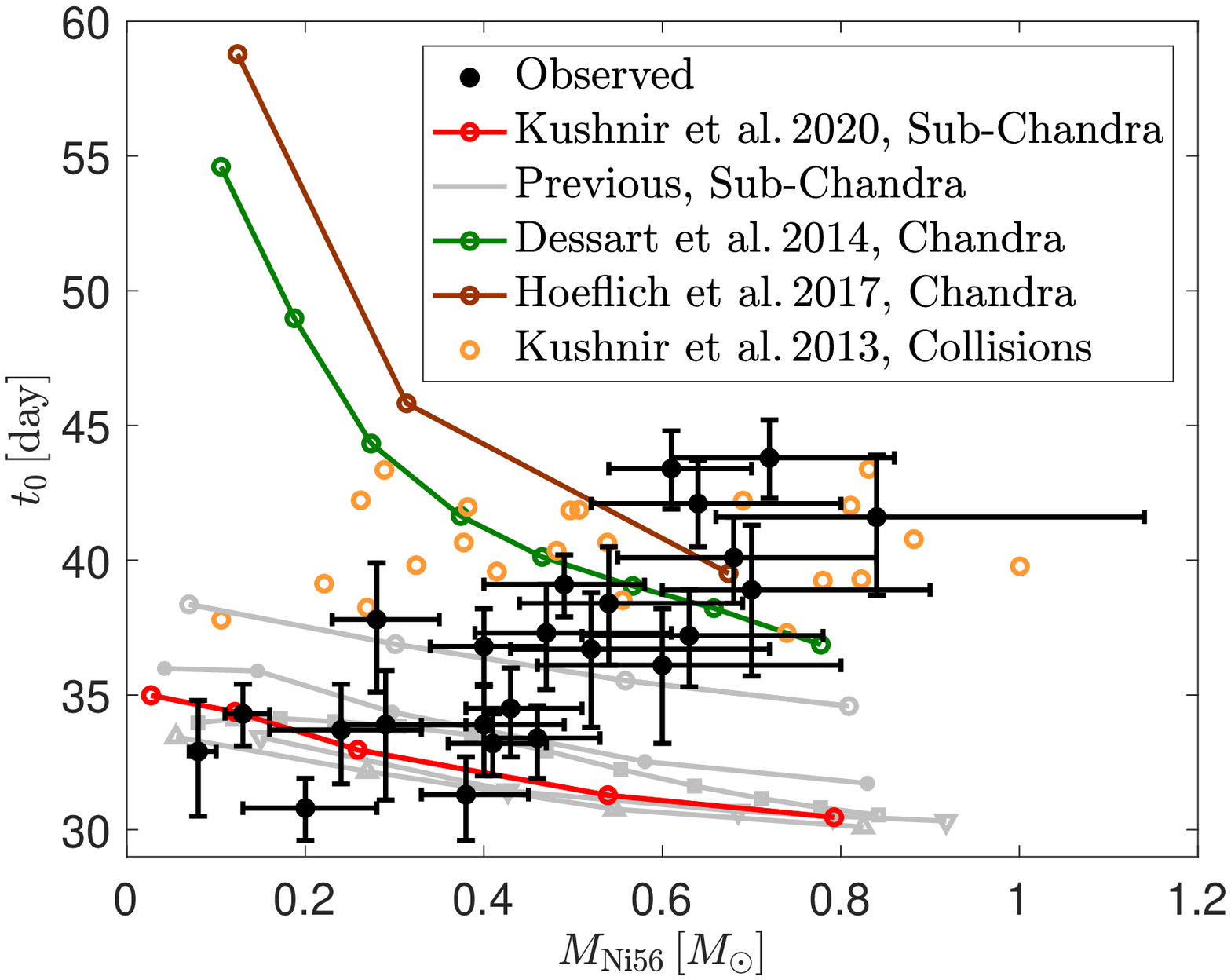}
\caption{The $t_0-\nimass$ relation. Black-filled circles: the observed SNe Ia sample of \citet[][]{Sharon2020}. Plotted are the median of the posterior distribution, together with the $68\%$ confidence levels. Red line: sCh results of \citet{Kushnir2020}. Grey lines: sCh results of previous studies \citep{Sim2010,Moll2014,Blondin2017,Shen2018,Bravo2019} are marked with circles, upward-pointing triangles, squares, filled circles and downward-pointing triangles, respectively. Green line: Ch results of \citet{Dessart2014}. Brown line: Ch results of \citet{Hoeflich2017}. Orange circles: The direct-collisions results of \citet{Kushnir2013}. \label{fig:1}}
\end{center}
\end{figure}

An observational property that can be calculated more robustly is the $\gamma$-ray escape time, $t_0$ \citep{Stritzinger2006,Scalzo2014,Wygoda2019a}, defined by \citep{Jeffery1999}
\begin{equation}\label{eq:dep_late}
f_\text{dep}(t) = \frac{t_0^2}{t^2},\;\;\;f_\text{dep}\ll 1,
\end{equation}
where $t$ is the time since explosion and $f_\text{dep}(t)$ is the $\gamma$-ray deposition function, which describes the fraction of the generated $\gamma$-ray energy that is deposited in the ejecta. For a small enough $\gamma$-ray optical depth, each $\gamma$-ray photon has a small chance of colliding with matter, such that the deposition function is proportional to the column density, which scales as $t^{-2}$. The value of $t_0$ can be measured from a bolometric light curve to a few percent accuracy \citep{Wygoda2019a,Sharon2020} due to an integral relation derived by \citet{Katz2013}, independent of the supernova distance. Together with $\nimass$, the $^{56}$Ni mass synthesized in the explosion, an observed $t_0-\nimass$ relation can be constructed \citep{Wygoda2019a}, see Figure~\ref{fig:1}. The accurate determination of $t_0$ by \citet{Sharon2020} reveals a positive correlation between $t_0$ and $\nimass$. The methods used in previous works did not allow a robust determination of such a correlation, although there were some hints for its existence \citep{Stritzinger2006,Scalzo2014,Wygoda2019a}. The observed $t_0-\nimass$ correlation is similar to the Phillips relation \citep{Phillips1993}, which relates the maximum flux to the width of the light curve in some band. However, unlike the Phillips relation, comparing to the $t_0-\nimass$ relation bypasses the need for radiation transfer calculations, as the value of $t_0$ can be directly inferred from the ejecta.

\citet{Wygoda2019a} showed that Ch models predict an anti-correlation between $t_0$ and $\nimass$ and that they deviate by almost a factor of two from the $t_0-\nimass$ relation for low-luminosity SNe Ia (see Figure~\ref{fig:1}). The Ch models cannot be saved for the low-luminosity events, unless the $^{56}$Ni is concentrated at very high velocities, which would be in direct conflict with the nebular spectra observations \citep{Kushnir2013,Wygoda2019a}. \citet{Kushnir2020} used a new burning scheme \citep{KK2019}, which consistently resolves the propagating burning wave, to calculate $\nimass$ and $t_0$ of one-dimensional sCh models with an accuracy of a few percent (see Figure~\ref{fig:1}). As can be seen in the figure, the sCh model predicts an anti-correlation between $t_0$ and $\nimass$, with $t_0\approx30\,\textrm{day}$ for luminous ($\nimass\gtrsim0.5\,M_{\odot}$) SNe Ia, while the observed $t_0$ is in the range of $35-45\,\textrm{day}$. Various physical and numerical uncertainties are unlikely to resolve the tension with observations, but they can deteriorate the agreement with observations for low-luminosity SNe Ia. The tension with the observed $t_0-\nimass$ relation exists in all previous studies of the sCh model (see Figure~\ref{fig:1}). \citet{Kushnir2020} argued that multi-dimensional modifications of the sCh model are unlikely to resolve the tension with observations. Also, the direct-collisions results of \citet{Kushnir2013} do not reproduce the observed $t_0-\nimass$ correlation, as they predict $t_0\approx40\,\textrm{day}$ for the entire $\nimass$ range (see Figure~\ref{fig:1}).

While all known models fail to reproduce the observed positive $t_0-\nimass$ correlation, there exists the possibility that two or more different progenitor scenarios conspire to produce a positive $t_0-\nimass$ correlation. For example, could it be that high-luminosity SNe are the result of Ch explosions while low-luminosity SNe are the result of sCh explosions? It is unlikely that the answer is positive. First, none of the models actually explains the region of $\nimass\approx0.5\,M_{\odot}$, where both should merge. Additionally, there is no apparent feature in any SNe Ia observation that hints at a change of progenitor around $\nimass\approx0.5\,M_{\odot}$, and if anything, the scatter around the Phillips relation in that region is so small \citep{Burns2018} that an extreme fine-tuning between the two models would be required for such a scenario to occur. Moreover, even for the range of luminosities where the models seem to explain the $t_0-\nimass$ relation, the models fail to explain some other robust observations. sCh models seem to be in conflict with the double-peaked or highly shifted $^{56}$Ni mass-weighted line-of-sight velocity distribution for a large fraction of low-luminosity events, as measured from nebular spectra observations \citep{Dong2015,Dong2018,Vallely2020}. Ch models predict a large mass ratio of Ni/Fe, which seems to be in conflict with nebular spectra observations of luminous SNe Ia \citep{Flors2020}, where the derived ratio is consistent with the lower predictions of sCh. 

While sCH calculations with the initial compositions suggested by evolutionary models of WDs are unable to explain the observed $t_0-\nimass$ relation, perhaps a different (heavier atoms) initial composition for high mass WDs would bring the predictions into better agreement with the observations (since the value of $t_0$ would increase with less available thermonuclear energy). Such a heavier initial composition is indeed expected for very massive, $\MWD\gtrsim1.1\,M_{\odot}$, WDs \citep[see, e.g.,][]{Lauffer2018}, however, the exact $\MWD$ for the transition as well as the exact initial composition (for all WD masses) are quite uncertain. We intend to study in a subsequent work whether there are initial compositions that can reproduce the $t_0-\nimass$ relation. Another possibility is that more accurate calculations of the direct-collision model would bring the predictions into better agreement with the observations, since the calculations of \citet{Kushnir2013} were with a resolution of $\textrm{few}\times\textrm{km}$ and with a simplified $13$-isotope reaction network, which are known to provide inaccurate results for low $\nimass$ values \citep{Moll2014,Kushnir2020}. We will report the results of more accurate calculations in a subsequent work.

\end{document}